\documentclass[%
reprint,
nofootinbib,
superscriptaddress,
showpacs,preprintnumbers,
amsmath,
amssymb,
aps,
prl
secnumarabic, graphics,floatfix,nofootinbib,
tightenlines,nobibnotes]{revtex4-1}

\usepackage{amsfonts}
\usepackage{cancel}
\usepackage{graphicx}
\usepackage{dcolumn}
\usepackage{bm}
\usepackage{hyperref}
\usepackage[mathlines]{lineno}
\usepackage{color}
\usepackage{braket}
\usepackage[normalem]{ulem}
\usepackage[english]{babel}
\definecolor{dgreen}{rgb}{0,0.39,0}
\usepackage{subfig}

\usepackage{color}

\newcommand{\diff}{\mathrm{d}}

\begin{document}
	\title{Minimal Length Effects on Quantum Cosmology and Quantum Black Hole Models}
	\author{Pasquale Bosso}
	\email{bosso@fisica.ugto.mx}
	\author{Octavio Obreg\'on}
	\email{octavio@fisica.ugto.mx}
	\affiliation{Departamento de F\'\i sica, Divisi\'on de Ciencias e Ingenier\'\i a,\protect\\ Universidad de Guanajuato, Campus Le\'on,\protect\\ Loma del Bosque 103, Frac. Lomas del Campestre, Le\'on, Gto., M\'exico}
	%

\begin{abstract}
A Kantowski-Sachs model with a modified quantization prescription is considered.
Such quantization rules, inspired by the so-called Generalized Uncertainty Principle (GUP), correspond to a modified commutation relation between minisuperspace variables and their conjugate momenta.
For a wide range of the modification parameter, this approach differentiates from the standard results by the presence of a potential well in the corresponding Wheeler-DeWitt equation.
This then produces the appearance of a set of wave functions, with corresponding discrete energy spectrum.
\end{abstract}

\pacs{04.60.--m, 04.60.Bc, 98.80.Qc}

\maketitle

\section{Introduction}

Among the most tenacious efforts in fundamental physics is the seek for a theory of Quantum Gravity (QG), that is, a theory that could give a quantum description of gravity.
As of now, several candidates have been proposed, and the debate on the validity of one over the others is open.
Despite the various theoretical possibilities to realize a theory of QG, no experimental evidence can direct us, nor any evidence of deviations from general relativity or quantum theories can help us in this task.
Nonetheless, there are a series of features that we expect from a quantum theory of gravity.
On of these features, when translated to low energy systems, consists in the existence of a minimal measurable length \cite{Garay1995_1}.
In fact, such a minimal length arises in different contexts, for example from string theory \cite{Gross1988_1,Amati1989_1}, loop quantum gravity \cite{Rovelli1995,Rovelli1998,Ashtekar2017}, and thought experiments in black hole physics \cite{Maggiore1993_1,Scardigli1999_1}.
These common characteristics of several theories of quantum gravity led to a phenomenological model consisting in a modification of the uncertainty principle.
Such a model is known as the generalized uncertainty principle (GUP).
It has been the subject of many studies, in the attempt to use it as a signature of QG, and has been compared with known phenomena and theories of modified gravity (see, \emph{e.g.}, \cite{Scardigli2016,Scardigli2016_1,Bosso2017,Lambiase2018,Bosso2018_1,Scardigli2018,Kanazawa2019,Buoninfante2019,Bushev2019}).
A version of this model considers a modification of the Heisenberg algebra \cite{Kempf1995_1,Das2009_1,BossoPhD,Bosso2018} to reproduce, via the Schr\"odinger--Robertson uncertainty relation, the desired minimal length.
Notice that this modification can also be thought of as a modified quantization rule.

These effects are predicted to be important in systems with energies near the Planck scale.
A particularly relevant example of such systems is the very early universe, in which quantum effects of gravity are expected to be dominant \cite{Ryan1972,Hartle1983,Ryan2015}.
A special branch of this line of investigation is loop quantum cosmology, developed in the past years, where the framework of loop quantum gravity has been applied to cosmology \cite{Ashtekar2011,Agullo2017}.
Therefore, Quantum Cosmology is the appropriate playground where this modified quantization rule is expected to be influential.
Previous approaches to this field using the tools of canonical quantum gravity have investigated various aspects of this construction with the purpose of studying quantum cosmological models.
In the past, several quantization procedures have been considered regarding this approximation (see, \emph{e.g.} \cite{Bergeron2017}).
In particular, recent attempts have been directed towards a noncommutative deformation of quantum cosmology \cite{Garcia-Compean2002,Aguero2007}, that is, descriptions in which variables do not commute.
This resulted from proposals of noncommutativity in spacetime and from developments in $M$ theory and string theory \cite{Snyder1947,Connes1995,Douglas2001}.

In the present work, thus, we implement a different perspective, proposing a quantization rule for the minisuperspace approximation \cite{Misner1969, Misner1972} in which the corresponding variables are considered to obey a similar commutation relation as in GUP.
This will imply a modification of the Wheeler-DeWitt equation (WDWE), governing the quantum cosmological model, characterizing a modified dynamics of the solution.
Previous approaches from a different viewpoint have been pursued in \cite{Kober2011,Faizal2014,Garattini2016}.
It is worth noticing that this procedure does not directly implies a physical minimal length.
Rather, it can be understood as imposing a minimal uncertainty in the minisuperspace variables.

As a particular case for this proposal, we will consider its effects on the Kantowski-Sachs model.
As it is known, at the classical level it describes a homogeneous but anisotropic cosmological model, thus not relatable with the current description with the observable universe \cite{Obregon1998}.
However, its relevance arises as well from the fact that it can describe a Schwarzschild black hole \cite{Gambini2015}.
The wave function of the corresponding quantum model thus represents a quantum cosmology or a quantum black hole.
The minisuperspace coordinates, at the present quantum stage, are not affected by their classical dependence on the time $t$ or the radius $r$ \cite{Obregon1998}.
Thus, from now on, we will refer our analysis only at the level of the minisuperspace Kantowski-Sachs variables and their quantum evolution.
This metric can be written as \cite{Misner1972}
\begin{multline}
\diff s^2 = - N^2 \diff t^2 + e^{2\sqrt{3} \beta} \diff r^2\\
+ e^{-2 \sqrt{3} \beta} e^{-2 \sqrt{3} \Omega} (\diff \theta^2 + \sin^2 \theta \diff \phi^2). \label{eqn:KS_metric}
\end{multline}
The corresponding WDWE in the standard theory of quantum gravity is given by
\begin{equation}
e^{\sqrt{3} \beta + 2 \sqrt{3} \Omega} \left[- P_\Omega^2 + P_\beta^2 - 48 e^{-2 \sqrt{3} \Omega} \right] \psi(\Omega,\beta) = 0, \label{eqn:WDWE_standard}
\end{equation}
where
\begin{align}
P_\Omega = & - i \frac{\partial}{\partial \Omega}, & P_\beta = & - i \frac{\partial }{\partial \beta},
\end{align}
are the conjugate momenta to the variables $\Omega$ and $\beta$, respectively, and such that
\begin{align}
[\Omega,P_\Omega] = & i, & [\beta,P_\Omega] = & 0, &
[\Omega,P_\beta] = & 0, & [\beta,P_\beta] = & i. \label{eqn:std_commutators}
\end{align}
The solutions of Eq. \eqref{eqn:WDWE_standard} are given in terms of the modified Bessel function $K_{i\nu}$ as follows
\begin{equation}
\psi^{\pm}_\nu (\Omega,\beta) = e^{\pm i \nu \sqrt{3} \beta} K_{i \nu} (4 e^{-\sqrt{4} \Omega}). \label{eqn:wf_KS}
\end{equation}

In what follows, we will revise the model above modifying the quantization relations in Eq. \eqref{eqn:std_commutators}.
In particular, we will consider a different commutation relation between the variables $\{\Omega,\beta\}$ and the conjugate momenta.
In fact, we will consider the model inspired by \cite{Kempf1995_1}, for which
\begin{equation}
[q_j,p_k] = i \delta_{jk} \{1 + \gamma^2 p_l p_l\}, \label{eqn:GUP}
\end{equation}
with
\begin{align}
q_1 = & \Omega, & q_2 = & \beta, & p_1 = & P_\Omega, & p_2 = & P_\beta,
\end{align}
where $\gamma$ is some parameter with units of inverse $P_\Omega$ and $P_\beta$, and where we considered Einstein's summation convention.
For a more convenient treatment, we will introduce coordinates $q'_j$ such that $[q'_j,p_k] = i \delta_{jk}$, \emph{i.e.} $q'_j$ and $p_k$ fulfill the same relations as those in Eqs. \eqref{eqn:std_commutators}.
The momentum-space representation of the coordinate operators obeying Eqs.~\eqref{eqn:GUP} is
\begin{equation}
q_j = i (1 + \gamma^2 p_k p_k) \frac{\partial}{\partial p_j} = (1 + \gamma^2 p_k p_k) q'_j.
\end{equation}
Notice that in this model, the two coordinates do not commute
\begin{equation}
[q_j,q_k] = 2 i \gamma^2 (1 + \gamma^2 p_l p_l) (p_j q'_k - p_k q'_j) = 2 \gamma^2 \epsilon_{jk} p_j q_k,
\end{equation}
where $\epsilon_{jk}$ is the two-indices Levi-Civita symbol.
%
Furthermore, in position-space we can write
\begin{equation}
q_j = q_j' (1 + \gamma^2 p_k p_k).
\end{equation}
As we will see next, this modification is directly related with the form of the wave function $\psi(\Omega,\beta)$ in Eq. \eqref{eqn:WDWE_standard} introducing, for a particular range of values for the parameter $\gamma$, a well in the potential in Eq. \eqref{eqn:WDWE_standard}.
The effect of this modification is to modify the uncertainty relation for the minisuperspace variables.
In fact, it imposes a minimal uncertainty in these variables, thus resulting in a fuzy metric.
As a consequence, this furthermore results in a notion of distance with a minimal uncertainty and, therefore, a minimal length.

This paper is structured as follows:
in Section \ref{sec:GUP}, we will revise the WDWE for the Kantowski-Sachs model with a modified quantization rule;
in Section \ref{sec:HO_approx}, we will focus on a particular region of the variables, in which the modified potential associated with the Kantowski-Sachs model produces a more noticeable difference with the usual quantum behavior Eq. \eqref{eqn:wf_KS};
finally, Section \ref{sec:Conclusion} is devoted to conclusions and outlook.

\section{Kantowski-Sachs model with GUP}\label{sec:GUP}

\begin{figure}
%
	\includegraphics[width=0.98\columnwidth]{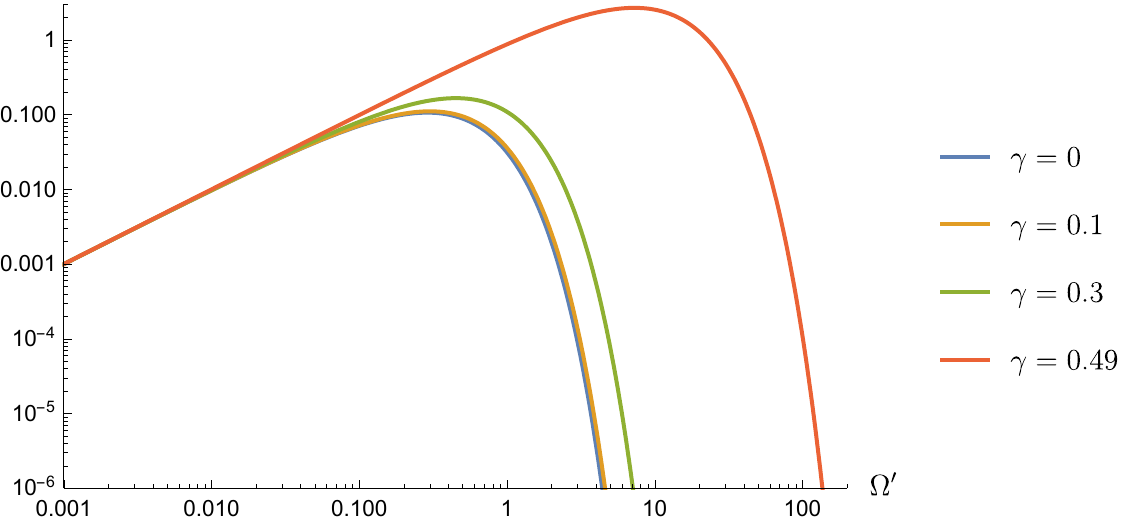}
	\caption{Functions $\Omega' e^{-2 \sqrt{3} (1 - 4 \gamma^2) \Omega'}$.
		Notice that the interval of values of $\Omega'$ in which the corrections are not negligible depends on the value of $\gamma$.
		In particular, the closer $\gamma$ is to the value $1/2$, the more extended and large in magnitude the modification is.
		The pick of this function is for $\Omega' = \frac{1}{2 \sqrt{3} (1 - 4 \gamma^2)}$.}
	\label{fig:Omega}
\end{figure}

Following \cite{Garcia-Compean2002} and using the relations above, the potential term in Eq. \eqref{eqn:WDWE_standard} can be rewritten as
\begin{multline}
	V = - 48 e^{-2 \sqrt{3}\Omega}
	= - 48 e^{- 2 \sqrt{3} \Omega' [1 + \gamma^2 (- P_\Omega^2 + P_\beta^2)]} \\
	\simeq
%
	- 48 
	e^{- 2 \sqrt{3} (1 - 4 \gamma^2) \Omega'}
	e^{- 2 \sqrt{3} \gamma^2 \Omega' (- P_\Omega^2 + P_\beta^2)}
	e^{12 i \gamma^2 \Omega' P_\Omega}, \label{eqn:potential}
\end{multline}
where Zassenhaus formula
\begin{equation}
	e^{A+B} = e^A
		e^B
		e^{- \frac{1}{2} [A,B]}
		e^{\frac{1}{6} ([A,[A,B]] + 2 [B,[A,B]])}
		\cdots\,,
\end{equation}
has been used and where only terms up to exponentials in $\gamma^2$ have been retained.
Using the substitution $\Omega' = e^x$ we find
\begin{equation}
	- i \Omega' \frac{\partial}{\partial \Omega'} = - i \frac{\partial}{\partial x}\,.
\end{equation}
Therefore, assuming the following representation for the momentum operators
\begin{align}
P_\Omega = & - i \frac{\partial}{\partial \Omega'}\,, & P_\beta = & - i \frac{\partial}{\partial \beta'}\,,
\end{align}
the last exponential above, Eq. \eqref{eqn:potential}, acts as a translation operator for the coordinate $x$, corresponding to a scaling of the coordinate $\Omega'$
\begin{equation}
	e^{12 i \gamma^2 \Omega' P_\Omega} \psi(\Omega',\beta)
	= \psi(e^{12 \gamma^2} \Omega',\beta).
\end{equation}
The potential above then becomes
\begin{multline}
	V \psi(\Omega',\beta') =
	- 48 e^{- 2 \sqrt{3} (1 - 4 \gamma^2) \Omega'} \\
	\times e^{- 2 \sqrt{3} \gamma^2 \Omega' (- P_\Omega^2 + P_\beta^2)} \psi(e^{12 \gamma^2} \Omega',\beta').
\end{multline}
We can further expand the second exponential in the previous expression up to second order in momentum, obtaining
\begin{equation}
	V \simeq
	- 48 e^{- 2 \sqrt{3} (1 - 4 \gamma^2) \Omega'}
	\left[1 - 2 \sqrt{3} \gamma^2 \Omega' (- P_\Omega^2 + P_\beta^2)\right].
\end{equation}
We can then rewrite the modified Eq. \eqref{eqn:WDWE_standard} as
\begin{multline}
	\left[\left(1 + 96 \sqrt{3} \gamma^2 \Omega' e^{- 2 \sqrt{3} (1 - 4 \gamma^2) \Omega'}\right) \left(- P_\Omega^2 + P_\beta^2\right) \right.\\
	\left.- 48 e^{- 2 \sqrt{3} (1 - 4 \gamma^2) \Omega'}\right] \psi(e^{12 \gamma^2} \Omega',\beta') = 0.
\end{multline}
It is interesting to notice that the region in which the modification terms are relevant depends on $\gamma$.
In fact, the closer $\gamma$ is to the value $1/2$, the more extended this region is and the more relevant the correction terms are, as shown in Fig. \ref{fig:Omega}.
We will focus on the interval of $\Omega'$ in which the modification is not negligible, since outside this region the same results as in \cite{Misner1972} apply.

Using the same factorization as in  \cite{Garcia-Compean2002},
\begin{equation}
	\psi(e^{12 \gamma^2} \Omega',\beta') = e^{\sqrt{3} \nu \beta'} \chi(\Omega'),
\end{equation}
we can write the previous equation up to second order in $\gamma$ as
\begin{multline}
	\left[\frac{\diff^2}{\diff \Omega' {}^2} - 3 \nu^2 - \left(1 - 96 \sqrt{3} \gamma^2 \Omega' e^{- 2 \sqrt{3} (1 - 4 \gamma^2) \Omega'}\right) \right.\\
	\left. \times 48 e^{- 2 \sqrt{3} (1 - 4 \gamma^2) \Omega'}\right] \chi(\Omega') \\
	= \left[- \frac{\diff^2}{\diff \Omega' {}^2} - V_{\gamma,\nu} - \gamma^2 \tilde{V}\right] \chi(\Omega') = 0, \label{eqn:WDW}
\end{multline}
where
\begin{subequations}
\begin{align}
	V_{\gamma,\nu} = & 3 \nu^2 + 48 e^{- 2 \sqrt{3} (1 - 4 \gamma^2) \Omega'},\\
	\tilde{V} = & - 4608 \sqrt{3} \Omega' e^{- 4 \sqrt{3} (1 - 4 \gamma^2) \Omega'}. \label{eqn:potentials}
\end{align}
\end{subequations}
The function $V_{\gamma,\nu}$, in the limit $\gamma \rightarrow 0$, represents the potential of the standard WDWE Eq. \eqref{eqn:WDWE_standard}.
On the other hand, $\tilde{V}$ represents the correction due to the modified commutation relation.
Notice that $\tilde{V}$ is relevant only in an interval about the value $\Omega' = \frac{1}{4 \sqrt{3} (1 - 4 \gamma^2)}$, whose extension depends on the value of $\gamma$. 
In what follows, thus, we will focus our attention around this value.
It is also interesting to notice that the correction does not depend on the parameter $\nu$.

\begin{figure}
	\subfloat[]{\includegraphics[width=\columnwidth]{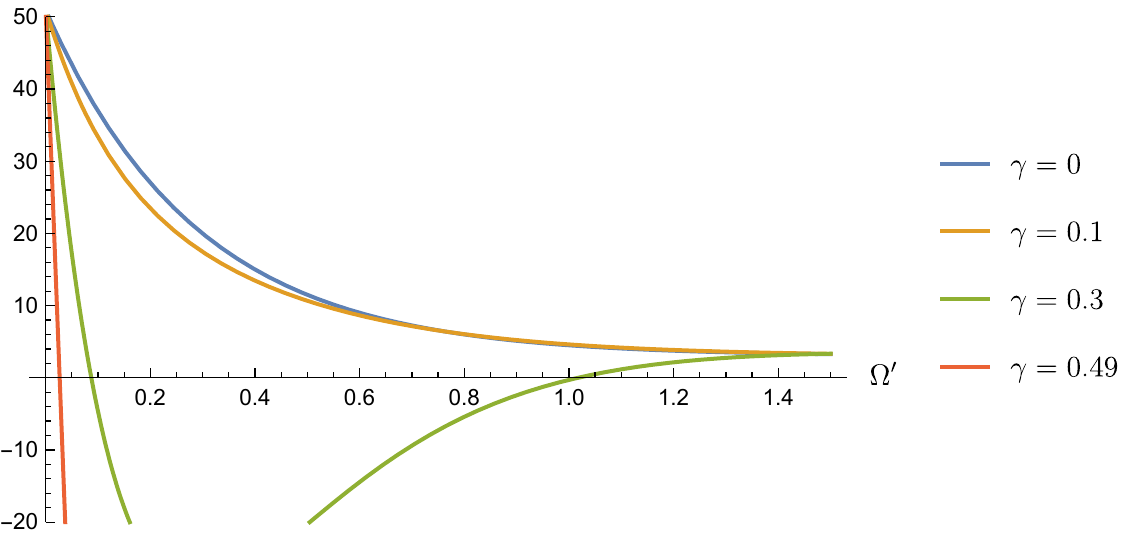}}
	
	\subfloat[]{\includegraphics[width=\columnwidth]{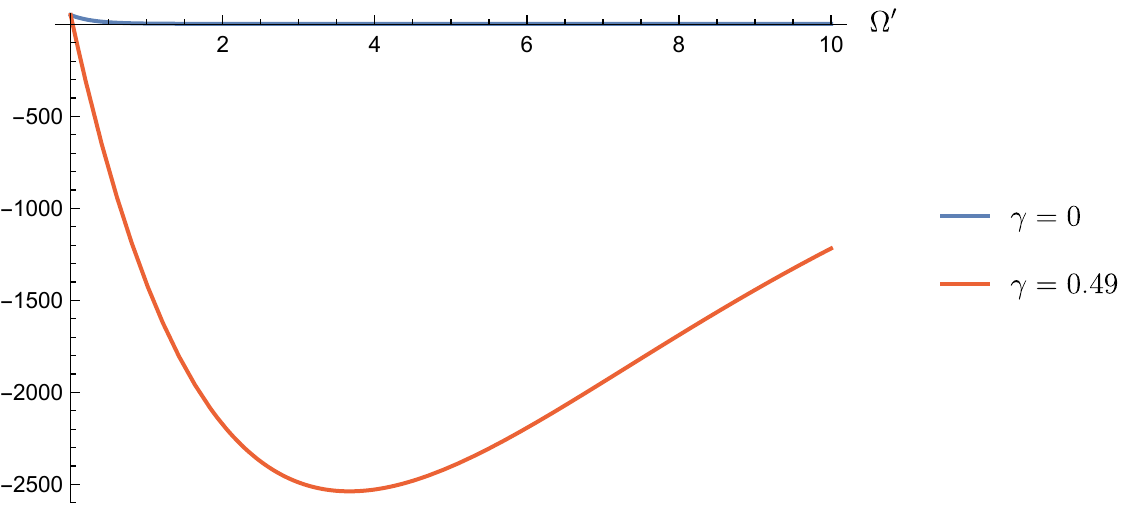}}
	\caption{Dependence of $V_{\gamma,\nu} + \gamma^2 \tilde{V}$ on $\gamma$.
		Four cases are reported, namely for the values $\gamma=0$, \emph{i.e.} the standard case, $\gamma=0.1$, $\gamma=0.3$, $\gamma=0.49$.
		This last case is shown also in figure b, compared with the standard case, for clarity.
		Notice the well for this last case, due to the predominance of $\gamma^2 \tilde{V}$ over $V_{\gamma,\nu}$.}
	\label{fig:potential}
\end{figure}

\begin{figure}
	\includegraphics[width=\columnwidth]{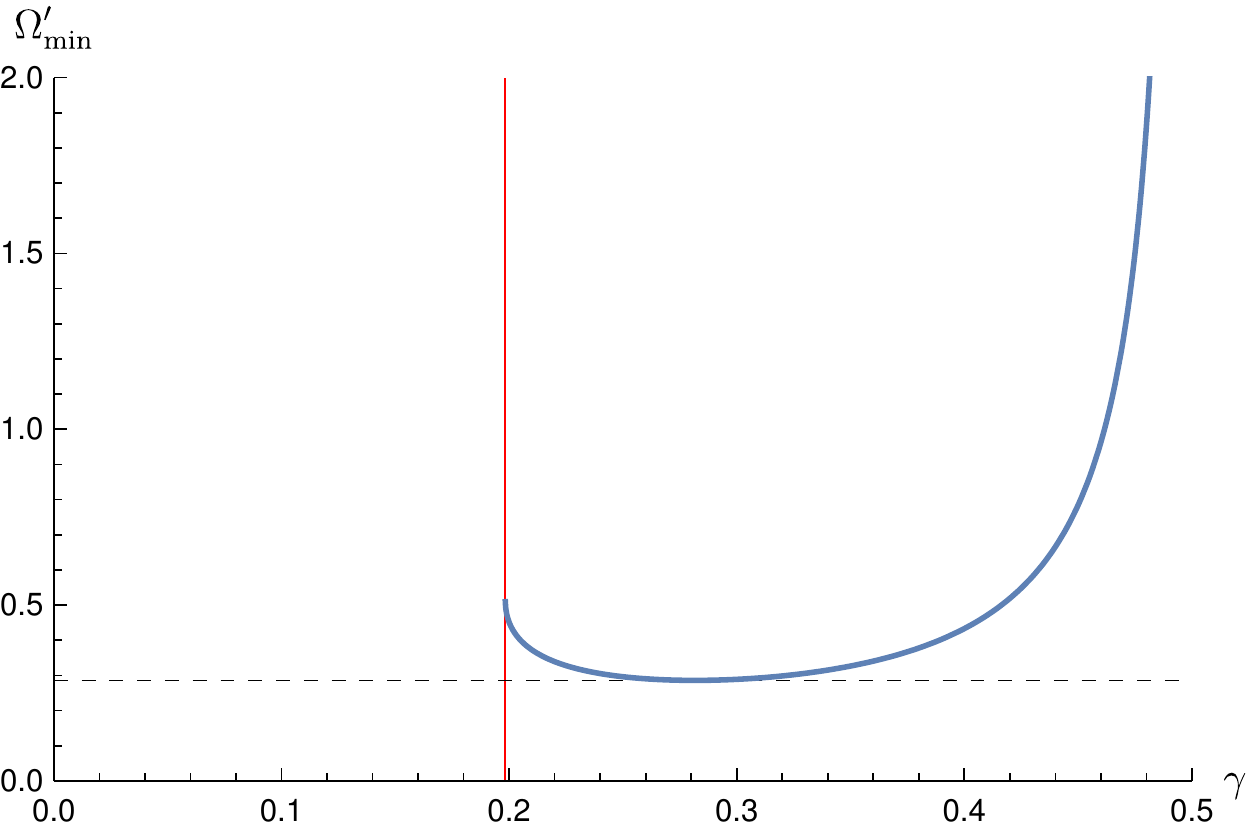}
	\caption{Value of $\Omega'_\mathrm{min}$ with respect to $\gamma$.
		The horizontal line, $\Omega_\mathrm{min} \simeq 0.28597$, corresponds to the smallest value for $\Omega_\mathrm{min}$.
		The vertical line correspond to the bound $\gamma = \frac{1}{2} \sqrt{\frac{e^{3/2}}{24 + e^{3/2}}}$, related with the appearance of the well in Eq. \eqref{eqn:WDW}.
		The asymptote at $\gamma = 1/2$ is due to the factor $1 - 4\gamma^2$ in Eq. \eqref{eqn:min_well}.}
	\label{fig:Omega_min}
\end{figure}

In this interval of values, the parameter $\gamma$ has a very interesting role.
In fact, for values $\gamma \ll \frac{1}{2} \sqrt{\frac{e^{3/2}}{24 + e^{3/2}}} \simeq 0.198$, the potential is mainly dominated by the standard part, $V_{\gamma,\nu}$.
On the other hand, for values $\gamma \geq \frac{1}{2} \sqrt{\frac{e^{3/2}}{24 + e^{3/2}}}$, the term $\tilde{V}$ dominates, introducing a well.
This is shown in Fig. \ref{fig:potential}.
For the same reason, the position of the local minimum of the potential shifts with $\gamma$.
It is given by the expression
\begin{equation}
	\Omega'_{\mathrm{min}} = \frac{\sqrt{3} \left[1 - 2 W\left(- \frac{\sqrt{e} \left(1 - 4 \gamma^2\right)}{96 \gamma^2}\right)\right]}{12 \left(1 - 4 \gamma^2\right)}, \label{eqn:min_well}
\end{equation}
where $W(z)$ is the Lambert $W$ function, solution of the equation $z = w e^w$ with respect to the variable $w$.
This function admits real values only for $z>-\frac{1}{e}$.
This motivates the bound $\gamma > \frac{1}{2} \sqrt{\frac{e^{3/2}}{24 + e^{3/2}}}$, as observed also in Fig. \ref{fig:Omega_min}.

\section{Harmonic Oscillator Approximation}\label{sec:HO_approx}

It is worth now to investigate further on the behavior of the solution of Eq. \eqref{eqn:WDW} in the well described above, that is for $\gamma > \frac{1}{2} \sqrt{\frac{e^{3/2}}{24 + e^{3/2}}}$.
To do this, let us consider an expansion of the the potential about $\Omega'_\mathrm{min}$ up to second order.
In this case, with the substitution $y = \Omega' - \Omega'_\mathrm{min}$, we find an equation that clearly resembles that of a harmonic oscillator
\begin{equation}
	\left\{-\frac{\diff^2}{\diff y^2} - b + a y^2 \right\} \chi(y) = 0, \label{eqn:HO_2order}
\end{equation}
with
\begin{subequations}
\begin{align}
	a = & 3 (1 - 4 \gamma^2)^3 \frac{\left[W \left(- \frac{\sqrt{e} (1 - 4 \gamma^2)}{96 \gamma^2}\right) + 1\right]}{\gamma^2 W^2 \left(- \frac{\sqrt{e} (1 - 4 \gamma^2)}{96 \gamma^2}\right)},\\
	b = & - 3 \left[\nu^2 - \frac{1 - 4 \gamma^2}{12 \gamma^2 W \left(- \frac{\sqrt{e} (1 - 4 \gamma^2)}{96 \gamma^2}\right)} \right.\\
	& \left. - \frac{1 - 4 \gamma^2}{24 \gamma^2 W^2 \left(- \frac{\sqrt{e} (1 - 4 \gamma^2)}{96 \gamma^2}\right)}\right],
\end{align}
\end{subequations}
where $b$ has the role of an energy.
Notice that this analogy is more appropriate the smaller $b$ is, as long as it is positive.
In other words, we require $b>0$ to obtain a bound state or, in terms of $\nu$,
\begin{equation}
	|\nu| < - \frac{\sqrt{(1 - 4 \gamma^2) \left[12 W \left(-\frac{\sqrt{e} (1 - 4 \gamma^2)}{96 \gamma^2}\right) + 6\right]}}{12 \gamma W \left(- \frac{\sqrt{e} (1 - 4 \gamma^2)}{96 \gamma^2}\right)}.
\end{equation}
Notice that the rhs is not necessarily real.
To obtain a real value for $\nu$, we need to impose the following further condition on $\gamma$
\begin{equation}
	\gamma \geq \frac{1}{2} \sqrt{\frac{e}{12 + e}}.
\end{equation}
Furthermore, this value for $\gamma$ is greater than the minimal value necessary to form a well in the potential.

\begin{figure}[b]
	\begin{center}
	\includegraphics[width=\columnwidth]{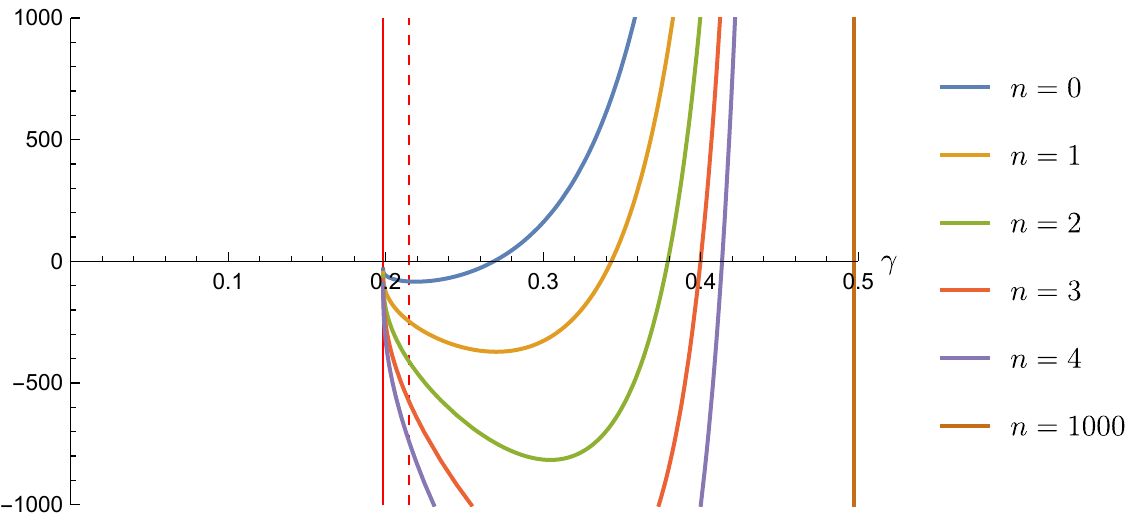}
	\end{center}
	\caption{Argument of the square root in Eq. \eqref{eqn:bound_states} for several values of $n$.
	The vertical line corresponds to the value $\gamma = \frac{1}{2} \sqrt{\frac{e^{3/2}}{24 + e^{3/2}}}$, while the dashed line corresponds to $\gamma = \frac{1}{2} \sqrt{\frac{e}{12 + e}}$.
	Notice that, for large $n$, the square root is real only for values $\gamma \rightarrow 1/2$.
	In other words, for values of $\gamma$ approaching 1/2, arbitrarily large numbers of bound states are allowed.}
	\label{fig:bound_states}
\end{figure}

Continuing in this analogy, and using the following redefinitions
\begin{align}
	E = & \frac{b}{2}, & \omega = & \sqrt{a}, \label{eqn:HO_energy}
\end{align}
and, furthermore, considering an harmonic oscillator with $\hbar = m = 1$, we have the following relation for the energy levels
\begin{equation}
	E = \omega \left(n + \frac{1}{2}\right) \qquad \Rightarrow \qquad b = \sqrt{a} (2 n + 1), \qquad n \in \mathbb{N}.
\end{equation}
This relation imposes a quantization rule for the parameter $\nu$ for a given value of $\gamma$
\begin{multline}
	\nu = \frac{\sqrt{1 - 4 \gamma^2}}{2 \sqrt{6}}
	\left\{\frac{2 W\left(- \frac{\sqrt{e} (1 - 4 \gamma^2)}{96 \gamma^2}\right) + 1}{\gamma^2 W^2\left(- \frac{\sqrt{e} (1 - 4 \gamma^2)}{96 \gamma^2}\right)} \right.\\
	\left.
		+ 8 \sqrt{3} (2 n + 1) \frac{\sqrt{(1 - 4 \gamma^2) \left[W\left(- \frac{\sqrt{e} (1 - 4 \gamma^2)}{96 \gamma^2}\right) + 1\right]}}{\gamma W\left(- \frac{\sqrt{e} (1 - 4 \gamma^2)}{96 \gamma^2}\right)}\right\}^{1/2}. \label{eqn:bound_states}
\end{multline}
Also in this case, looking for real values of $\nu$ gives constraints on $\gamma$ and the number of possible bounds states, as seen in Fig. \ref{fig:bound_states}.
Numerically, one finds that a first bound state is allowed for $\gamma > 0.268593$, two bound states appear when $\gamma > 0.343239$, three for $\gamma > 0.379114$, and so forth.
In general, a larger number of bound states are allowed for larger values of $\gamma$, provided that $\gamma < 1/2$.
In the limit $\gamma \rightarrow 1/2$, an infinite ladder of bound states is present.

\subsection*{Perturbing the Approximation}

For a better study of the effects of the proposed quantization rule, we will retain terms up to fourth order in $\Omega'$ in Eq. \eqref{eqn:WDW}.
Using the same substitution above, we can write
\begin{equation}
	\left\{-\frac{\diff^2}{\diff y^2} - b + a y^2 + c y^3 + d y^4\right\} \chi(y) = 0, \label{eqn:HO_4order}
\end{equation}
with
\begin{subequations}
\begin{align}
	c = & - 2 \sqrt{3} (1 - 4 \gamma^2)^4 \frac{\left[3 W \left(- \frac{\sqrt{e} (1 - 4 \gamma^2)}{96 \gamma^2}\right) + 4\right]}{\gamma^2 W^2 \left(- \frac{\sqrt{e} (1 - 4 \gamma^2)}{96 \gamma^2}\right)},\\
	d = & 3 (1 - 4 \gamma^2)^5 \frac{\left[7 W \left(- \frac{\sqrt{e} (1 - 4 \gamma^2)}{96 \gamma^2}\right) + 12\right]}{\gamma^2 W^2 \left(- \frac{\sqrt{e} (1 - 4 \gamma^2)}{96 \gamma^2}\right)}.
\end{align}
\end{subequations}
When these extra terms are small compared to the one already analyzed, one can use perturbation theory to compute the correction to the energy levels.

\begin{figure}
	\begin{center}
	\includegraphics[width=\columnwidth]{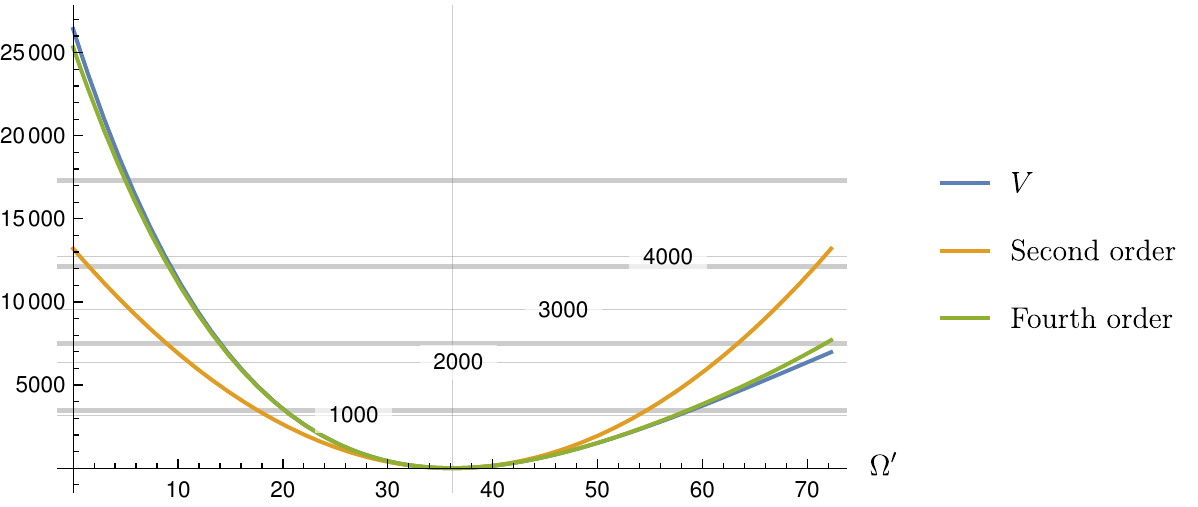}
	\end{center}
	\caption{Energies and corresponding corrections to bound states described by Eqs. \eqref{eqn:HO_2order} and \eqref{eqn:HO_4order} for $\gamma = 0.449$.
	Horizontal lines describe the energy levels for 4 particular values of the occupation number (indicated on the plot.)
	Thin lines correspond to the energy as computed via Eq. \eqref{eqn:HO_energy}, while thicker lines are the corrected energy, considering the term in Eq. \eqref{eqn:HO_corrections}.
	Finally, in the plot, the blue line corresponds to the full potential in Eq. \eqref{eqn:WDW}, the orange line correspond to the second order expansion in $\Omega'$ about $\Omega'_\mathrm{min}$, and the green one is the fourth order expansion.}\label{fig:corrected_energy}
\end{figure}

In general, when $n$ bound states are allowed, the energy of the $n$-th state will be corrected by a term
\begin{equation}
	\epsilon_{n,(1)} = \frac{6 n^2 + 6 n + 3}{4} (1 - 4 \gamma^2)^2 \frac{7 W \left(- \frac{\sqrt{e} (1 - 4 \gamma^2)}{96 \gamma^2}\right) + 12}{W \left(- \frac{\sqrt{e} (1 - 4 \gamma^2)}{96 \gamma^2}\right) + 1}. \label{eqn:HO_corrections}
\end{equation}
Notice that, for $\sqrt{\frac{e^{3/2}}{24 + e^{3/2}}} < 2 \gamma < 1$, these corrections are always positive and their magnitude increase quadratically with the occupation number $n$, as shown in Fig. \ref{fig:corrected_energy}.

Moreover, in general, if $n$ bound states are allowed, the correction to the $m$-th state is
\begin{widetext}
\begin{multline}
	|\eta_{m,(1)}\rangle
	= \sum_{\substack{s=0\\s\neq m}}^n \frac{\langle s|c x^3 + d x^4|m \rangle}{E_m - E_s}
	= \sum_{u=-\left\lfloor\frac{m+1}{2}\right\rfloor}^{\left\lfloor\frac{n-m-1}{2}\right\rfloor} c \frac{\langle m+2u+1|x^3|m \rangle}{E_m - E_{m+2u+1}} |m+2u+1\rangle
	+ \sum_{\substack{v=-\left\lfloor\frac{m}{2}\right\rfloor\\v\neq0}}^{\left\lfloor\frac{n-m}{2}\right\rfloor} d \frac{\langle m+2v|x^4|m \rangle}{E_m - E_{m+2v}} |m+2v\rangle \\
	= (1 - 4 \gamma^2)^{3/4} \frac{\left[3 W \left(- \frac{\sqrt{e} (1 - 4 \gamma^2)}{96 \gamma^2}\right) + 4\right] \sqrt{\gamma W \left(- \frac{\sqrt{e} (1 - 4 \gamma^2)}{96 \gamma^2}\right)}}{3^{3/4} \sqrt{2} \left[W \left(- \frac{\sqrt{e} (1 - 4 \gamma^2)}{96 \gamma^2}\right) + 1\right]^{5/4}} \\
	\times \left[ - \frac{\sqrt{m^{\underline{3}}}}{3} |m-3\rangle - 3 m^{3/2} |m-1\rangle + 3 (m+1)^{3/2} |m+1\rangle + \frac{\sqrt{(m+1)^{\overline{3}}}}{3} |m+3\rangle\right]\\
	- (1 - 4 \gamma^2)^2 \frac{\left[7 W \left(- \frac{\sqrt{e} (1 - 4 \gamma^2)}{96 \gamma^2}\right) + 12\right] \gamma W\left(- \frac{\sqrt{e} (1 - 4 \gamma^2)}{96 \gamma^2}\right)}{8 \sqrt{3} \left[W \left(- \frac{\sqrt{e} (1 - 4 \gamma^2)}{96 \gamma^2}\right) + 1\right]^{3/2}} \\
	\times \left[- \frac{\sqrt{m^{\underline{4}}}}{2} |m-4\rangle - 2 \sqrt{m^{\underline{2}}}(2m-2+1) |m-2\rangle + 2\sqrt{(m+1)^{\overline{2}}} |m+2\rangle + \frac{\sqrt{(m+1)^{\overline{4}}}}{2} |m+4\rangle\right].
\end{multline}
\end{widetext}

\section{Conclusion and Outlook} \label{sec:Conclusion}


Summarizing what has been found in this work, we have considered the Kantowski-Sachs model in the context of quantum cosmology with a modified quantization rule.
In doing so, one of the most interesting results is that, for $\gamma \ll 1/2$ but $\gamma \geq \frac{1}{2} \sqrt{\frac{e^{3/2}}{24 + e^{3/2}}}$, this modification has a deep impact only on a relatively restricted region of the coordinate space.
Furthermore, it is interesting to observe that this region, for a wide range of values of $\gamma$, is very close to or includes the most probable value for the variable $\Omega$ as found in \cite{Garcia-Compean2002}.
Therefore, it has a concrete influence on the considered model.
Furthermore, we have noticed that, for a particular interval of the modification parameter, a well appears in the quantum potential characterizing the system.
The presence of this well is a completely novel aspect of the application of this modification with respect to the standard quantum analysis of the Kantowski-Sachs minisuperspace model.
Because of this feature, the solution in that particular region and for given values of the parameter $\gamma$ can be expressed in terms of harmonic oscillator states, the number of which depends on $\gamma$ itself.

The importance of these results goes well beyond the cosmological aspects of Kantowski-Sachs model.
In fact, as mentioned above, this model would represent a possible quantum description of a spherically symmetric black hole \cite{Gambini2015}.
Therefore, continuing the works in \cite{Obregon2001,Arraut2009,Bargueno2015}, it would be possible to use the results presented in this paper to further study these effects on quantum black hole models.
In particular, the application of GUP in this context results in a minimal uncertainty for $\Omega$.
In turn, it would result in a minimal uncertainty for the radial coordinate of the black hole.
This and further analyses will be pursued in future works.

\section*{Acknowledgments}
O. Obreg\'on was supported by CONACYT Project 257919, UG Projects and PRODEP.
P. Bosso was supported by a PRODEP postdoctoral grant.




\end{document}